# Harmonic analysis of spherical sampling in diffusion MRI

A. Daducci[1], J. McEwen[2], D. Van De Ville[3,4], J-P. Thiran[1], and Y. Wiaux[2,4]

[1]Signal Processing Laboratory (LTS5), École Polytechnique Fédérale de Lausanne (EPFL), Lausanne, Switzerland, [2]Institute of Electrical Engineering, École Polytechnique Fédérale de Lausanne (EPFL), Lausanne, Switzerland, [3]Institute of Bioengineering, École Polytechnique Fédérale de Lausanne (EPFL), Lausanne, Switzerland, [4]Department of Radiology and Medical Informatics, University of Geneva (UniGE), Geneva, Switzerland

**Introduction.** In the last decade diffusion MRI has become a powerful tool to non-invasively study white-matter integrity in the brain. Recently many research groups have focused their attention on multi-shell spherical acquisitions [1, 2, 3] with the aim of effectively mapping the diffusion signal with a lower number of q-space samples, hence enabling a crucial reduction of acquisition time. One of the quantities commonly studied in this context is the so-called orientation distribution function (ODF). In this setting, the spherical harmonic (SH) transform has gained a great deal of popularity thanks to its ability to perform convolution operations efficiently and accurately, such as the Funk-Radon transform notably required for ODF computation from q-space data. However, if the q-space signal is described with an unsuitable angular resolution at any b-value probed, aliasing (or interpolation) artifacts are unavoidably created. So far this aspect has been tackled empirically and, to our knowledge, no study has addressed this problem in a quantitative approach. The aim of the present work is to study more theoretically the efficiency of multi-shell spherical sampling in diffusion MRI, in order to gain understanding in HYDI-like approaches [4], possibly paving the way to further optimization strategies.

**Theory.** The ODF is a function on the sphere defined as the radial integration of the full diffusion propagator P: $O(\hat{r}) = \int P(\hat{r}, r) r^2 dr$. The spherical harmonics $Y_{\ell m}(\hat{r})$, with $\ell \in \mathbb{N}$ and $|m| \leq \ell$, represent an orthonormal basis for functions on the sphere. We have proved the following original expression for the SH coefficients of the ODF as a function of the q-space signal E, the 3D Fourier transform of P (see [3] for a similar expression in function space): $O_{\ell m} = \frac{\delta_{\ell 0} \delta_{m 0}}{\sqrt{4\pi}} + \frac{\ell(\ell+1)}{4\pi} P_\ell(0) S_{\ell m}$, where $S_{\ell m}$ represent the SH coefficients of the following radial integration of the q-space signal E, $S(\hat{q}) = \int \frac{E(\hat{q}, q) - 1}{q} dq$, with $\ell < B_0$, where $B_0$ represents the angular band limit of the ODF. As acknowledged previously (see [5]), the Funk-Radon transform is represented efficiently in harmonic space by the simple factor $P_\ell(0)$. However, to our knowledge all research in diffusion MRI is based on spherical sampling distributions providing a non-theoretically exact SH transform. In this work, we advocate the use of equiangular grids, on which sampling theorems exist [6, 7] that allow the exact computation of the $B^2$ SH coefficients of a function on the sphere of band limit B on the basis of no more than exactly $2B^2 - 3B + 2$ samples [7], and by means of fast algorithms.

**Methods.** To simulate the diffusion process occurring in human brains, synthetic data were generated by means of the multi-tensor model [8] using diffusivity values typically observed in real cases ($\lambda_1 \in [1, 2] \cdot 10^{-3}$ mm$^2$/s and $\lambda_2 = \lambda_3 \in [0.1, 0.6] \cdot 10^{-3}$ mm$^2$/s, see [1, 5]). A two-fiber model was considered, with different values of crossing angle α and volume fractions $\rho_1$ and $\rho_2$, with $\alpha \in \{15°, 40°, 65°, 90°\}$ and $\rho_1 \in \{1, 0.75, 0.5\}$, and for various fractional anisotropies (FA) (identical for the two fibers), with FA $\in \{0.8, 0.6, 0.4\}$. Firstly, the value of the angular band limits B of a spherical cut, at given b-value, of the signal E was studied. The discrepancy between the modeled signal E at each b-value and its forward-inverse SH transform Ẽ was evaluated in terms of the sum of the square values of E-Ẽ. Given the exact sampling theorem available on equiangular grids, such a measure quantifies the error arising from any underestimation of the band limit B and leading to *angular aliasing*. Secondly, in order to illustrate this angular aliasing effect, ODFs were reconstructed from simulated multi-shell data acquisitions with B on each sampled shell associated with either a low or high aliasing level. A 45° crossing angle was considered, with $\rho_1 = 0.5$ and FA = 0.8, both without noise and with Rician noise at an SNR of 25. A multi-shell setting with 3 shells linearly sampled at b-values 1000, 3500 and 8000 s/mm$^2$ and a single-shell setting with 1 shell at b-value 3500 s/mm$^2$ were considered in order to illustrate the complementary *radial aliasing* induced by excessive reduction of the number of shells, in the fully model independent approach of interest. As a reference, in addition to the ground truth ODF computed analytically from the model, experiments were also simulated in a standard Diffusion Spectrum Imaging (DSI) setting using a state-of-the-art model independent approach.

**Results.** Figure 1 (on the right) reports the values of B required to describe the diffusion signal E at each b-value for the two angular aliasing levels and each FA considered. These values range between 1 and 11 and are naturally independent of the volume fractions. For a given level of aliasing, the graphs quantify two expected behaviors. Firstly, for a given FA, the required angular resolution increases with the b-value. Secondly, for a given b-value, the required angular resolution increases with FA.

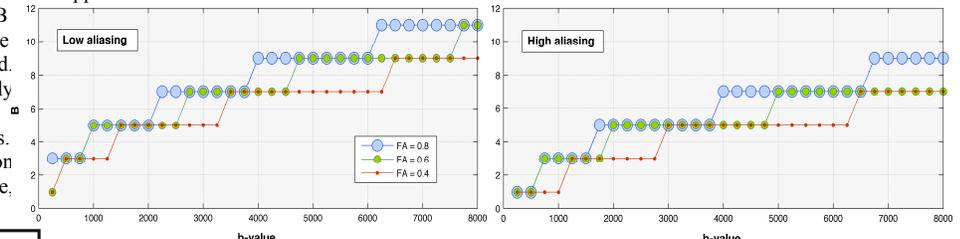

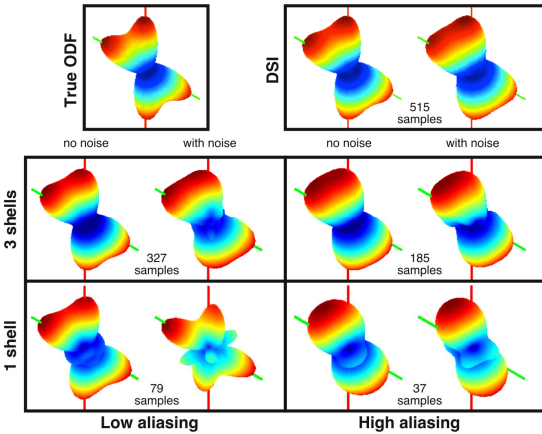

Figure 2 (on the left) shows the reconstructed 45° two-fiber ODFs from the multi-shell and single-shell experiments at the two angular aliasing levels considered. The multi-shell acquisitions with low angular aliasing required 327 samples to produce ODFs of equivalent quality to DSI, which is based on 515 Cartesian samples. Drastically reducing the number of samples, or equivalently allowing increased angular aliasing or increased radial aliasing in the 1-shell setting, lead to significant artifacts, potentially hampering the detection of the two fiber directions from the ODF.

**Conclusion.** This study presents a theoretical framework and numerical simulations aiming to provide initial guidance in designing efficient multi-shell spherical sampling strategies for diffusion MRI, in a model independent approach. Beyond the spherical sampling on equiangular grids advocated by the present study, the radial sampling might naturally be optimized either by efficient interpolation schemes or by model dependent approaches.